\documentclass[12pt,thmsa]{article}
\def \eq {\begin{equation}}
\def \fim-eq {\end{equation}}
\begin{document}

\title{A Cavity QED Implementation of Deutsch-Jozsa Algorithm }
\author{E. S. Guerra \\
%EndAName
Departamento de F\'{\i}sica \\
Universidade Federal Rural do Rio de Janeiro \\
Cx. Postal 23851, 23890-000 Serop\'edica, RJ, Brazil \\
email: emerson@ufrrj.br}
\maketitle

\begin{abstract}
The Deutsch-Jozsa algorithm is a generalization of the Deutsch algorithm
which was the first algorithm written. We present schemes to implement the
Deutsch algorithm and the Deutsch-Jozsa algorithm via cavity QED.

\end{abstract}

\section{INTRODUCTION}

We can say that quantum computation is a young and promising subject and can
revolutionize our computational way of tackling problems, especially the
simulation of quantum problems numerically. Nowadays quantum computation is
a branch of the wider subject \ quantum information \cite{Nielsen, MathQC,
PhysQI, QCAlgs} which embraces besides quantum computation, for instance,
quantum communication and quantum cryptography. \ P. Benioff \cite{Benioff}
and R. Feynmann \cite{Feynmann} were the pioneers \ of quantum computation
suggesting the build up of computers \ based on the principles of quantum
mechanics. Following the lead of Benioff and Feynmann, D. Deutsch in 1985 
\cite{Deutsch} made concrete proposals exploring some properties of quantum
mechanics to obtain unprecedented parallelism in computation which
represented a really breakthrough to the subject which since then has
developed quickly although we can still say that perhaps we are far from
building a working practical and economically viable hardware based on
quantum mechanics, that is, unfortunately we are still far from having, for
instance, a quantum PC \ for helping us in working out problems and, why
not, for our enjoyment. However, we should point out that although it seems
that a quantum computer would speed up the solution of some problems and
even turn out possible the solution of some problems intractable by
classical computation, nobody still knows what is the real power of a
quantum computer when compared with classical computers. Despite this fact
we should go on in the enterprise of developing research on this field which
can at least give us new insights on quantum theory and quantum information
science.

Important quantum algorithms are the Deutsch algorithm \cite{Deutsch, QCAlgs}%
, the Deutsch-Jozsa algorithm \cite{DeutschJozsa, QCAlgs}, the Simon
algorithm \cite{Simon, QCAlgs}, the Shor algorithm \cite{Shor, QCAlgs} and
the Grover algorithm \cite{Grover, QCAlgs}.

In this work we present an implementation of the Deutsch \cite{Deutsch,
QCAlgs}\ and the Deutsch-Jozsa algorithms \cite{DeutschJozsa, QCAlgs}, via
cavity QED. The Deutsch algorithm was the first concrete proposal of
computation making use of the special features of quantum mechanics. A
recent alternative proposal of realization of the Deutsch algorithm is
presented in \cite{Zheng}.

\section{CAVITY QED REALIZATION\protect\bigskip}

Let us start revising the Deutsch problem. Consider an arbitrary Boolean
function $F:\{0,1\}\longrightarrow \{0,1\}$. There are four such a functions
corresponding to two possible arguments and two possible values. For two of
them $F(0)=F(1)$ and in this case we say that $F$ is constant. For the cases
in which $F(0)\neq F(1)$ we say that $F$ is balanced. Suppose we do not know
the function and we are given an Oracle which can evaluate it and gives us
the result. Notice that in order to decide if $F$ is constant or balanced we
will have to use the oracle twice to know its value for 0 and 1. The Deutsch
algorithm can solve this problem with just one call of the oracle. Let us
see how the Deutsch algorithm works. First let us suppose we have a gate ($F-
$gate). The action of the $F-$gate is%
\begin{equation}
|x,y\rangle \longrightarrow |x,y\oplus F(x)\rangle .
\end{equation}%
The Deutsch algorithm employs a $F-$gate with $F$ being our function which
we want to decide if is constant or balanced. \ Considering that the input
two qubit state is%
\begin{equation}
|\psi _{in}\rangle =|x\rangle \frac{1}{\sqrt{2}}(|0\rangle -|1\rangle )=%
\frac{1}{\sqrt{2}}(|x,0\rangle -|x,1\rangle ),
\end{equation}%
the output is%
\begin{equation}
|\psi _{out}\rangle =\frac{1}{\sqrt{2}}(|x,F(x)\rangle -|x,1\oplus
F(x)\rangle ).
\end{equation}%
Since $f(x)=0$ or $1$ this can be written as 
\begin{equation}
|\psi _{out}\rangle =(-1)^{F(x)}\frac{1}{\sqrt{2}}(|x,0\rangle -|x,1\rangle
)=(-1)^{F(x)}|x\rangle \frac{1}{\sqrt{2}}(|0\rangle -|1\rangle ),
\end{equation}%
and as we see the effect is to change the state of the $|x\rangle $ qubit to 
$(-1)^{F(x)}|x\rangle .$ Therefore the value of the function is in the phase
of the state $|x\rangle $.

The actual circuit of the Deutsch algorithm is shown in Fig. 1 where $H$ is
a Hadamard gate%
\begin{equation}
H=\frac{1}{\sqrt{2}}\left[ 
\begin{array}{cc}
1 & 1 \\ 
1 & -1%
\end{array}%
\right]  \label{H}
\end{equation}%
and%
\begin{eqnarray*}
|0\rangle &\longrightarrow &\frac{1}{\sqrt{2}}(|0\rangle +|1\rangle ), \\
|1\rangle &\longrightarrow &\frac{1}{\sqrt{2}}(|0\rangle -|1\rangle ).
\end{eqnarray*}%
Then, as we see, after the action of the $F-$gate we have%
\begin{equation}
|0,1\rangle \longrightarrow \frac{1}{\sqrt{2}}[(-1)^{F(0)}|0\rangle
+(-1)^{F(1)}|1\rangle ]\frac{1}{\sqrt{2}}(|0\rangle -|1\rangle )
\end{equation}%
and, for $F(0)=F(1)$, the upper qubit is therefore in the state $\pm \frac{1%
}{\sqrt{2}}(|0\rangle +|1\rangle )$ and after the application of the last
Hadamard gate it will be in the state $\pm |0\rangle $. For $F(0)\neq F(1)$,
the upper qubit is therefore in the state $\pm \frac{1}{\sqrt{2}}(|0\rangle
-|1\rangle )$ and after the application of the last Hadamard gate it will be
in the state $\pm |1\rangle $. Thus a single call to the quantum oracle
followed by a measure of the upper qubit in the computational basis gives us
the answer of the problem. Notice that the state in which the lower qubit is
left is not important.

Let us see now how we can implement experimentally the Deutsch algorithm. We
start assuming that we have a cavity $C$ prepared in the state 
\begin{equation}
|-\rangle _{C}=\frac{(|0\rangle _{C}-|1\rangle _{C})}{\sqrt{2}}.  \label{C-}
\end{equation}%
In order to prepare this state, we send a two-level atom $A0$, with $%
|f_{0}\rangle $ and $|e_{0}\rangle $ being the lower and upper level
respectively, in the state 
\begin{equation}
\mid \psi \rangle _{A0}=\frac{1}{\sqrt{2}}(-i\mid e_{0}\rangle +\mid
f_{0}\rangle ),
\end{equation}%
through $C$, for $A0$ resonant with the cavity. If $g$ is the coupling
constant and $\tau $ the atom-field interaction time, under the
Jaynes-Cummings dynamics \cite{Orszag}, for $g\tau =\pi /2$, \ we know that
the state $|f_{0}\rangle |0\rangle _{C}$ does not evolve, however, the state 
$|e_{0}\rangle |0\rangle _{C}$ evolves to $-i|f_{0}\rangle |1\rangle _{C}$.
Then, for the cavity initially in the vacuum state $|0\rangle _{C}$, we have%
\begin{equation}
\frac{(|f_{0}\rangle -i|e_{0}\rangle )}{\sqrt{2}}|0\rangle
_{C}\longrightarrow |f_{0}\rangle \frac{(|0\rangle _{C}-|1\rangle _{C})}{%
\sqrt{2}}=|f_{0}\rangle |-\rangle _{C}
\end{equation}

Then let us assume that we have a cavity $C$ prepared initially in a state $%
\mid -\rangle _{C}$ and an atom $A1$ \ prepared initially the state

\begin{equation}
\mid \psi \rangle _{A1}=\frac{1}{\sqrt{2}}(\mid e_{1}\rangle +\mid
f_{1}\rangle ),  \label{Psiat}
\end{equation}%
Consider now that for an atom $Ak,$ with $|f_{k}\rangle $ and $|e_{k}\rangle 
$ being the lower and upper level respectively, the transition $\mid
f_{k}\rangle \rightleftharpoons \mid e_{k}\rangle $ is far from resonance
with the cavity central frequency as shown in Fig. 2. In this case the
effective Hamiltonian is given by \cite{EFFH}

\begin{equation}
H=\hbar \frac{g^{2}}{\Delta }(a^{\dagger }a+1)(\mid e_{k}\rangle \langle
e_{k}\mid -\hbar \frac{g^{2}}{\Delta }a^{\dagger }a\mid f_{k}\rangle \langle
f_{k}\mid ),  \label{H1}
\end{equation}%
and the time evolution operator is given by \cite{Orszag}

\begin{equation}
U(t)=e^{-i\varphi (a^{\dagger }a+1)}\mid e_{k}\rangle \langle e_{k}\mid
+e^{i\varphi a^{\dagger }a}\mid f_{k}\rangle \langle f_{k}\mid ,  \label{U1}
\end{equation}%
where $\varphi =g^{2}\tau /$ $\Delta $, $g$ is the coupling constant, $%
\Delta =\omega _{e}-\omega _{f}-\omega $ is the detuning where $\omega _{e}$
and $\omega _{f}$ \ are the frequencies of the upper and lower levels
respectively and $\omega $ is the cavity field frequency and $\tau $ is the
atom-field interaction time. Now, we \ are going to send $A1$ through cavity 
$C$ where the atom interacts dispersively with $C$ according to (\ref{U1}).
Let us take $\varphi =\pi .$ Then, after $A1$ flies through $C$ prepared in
state (\ref{C-}), we have%
\begin{equation}
|\psi \rangle _{A1-C}=\frac{1}{2}(-|e_{1}\rangle +|f_{1}\rangle )(|0\rangle
_{C}+|1\rangle _{C}).  \label{PsiA1Cbal}
\end{equation}%
If $\varphi =2\pi $ we have 
\begin{equation}
|\psi \rangle _{A1-C}=\frac{1}{2}(|e_{1}\rangle +|f_{1}\rangle )(|0\rangle
_{C}-|1\rangle _{C}).  \label{PsiA1Cconst}
\end{equation}%
Now, if we use the notation

\begin{eqnarray}
&\mid &e_{k}\rangle =\mid 0\rangle _{Ak},  \nonumber \\
&\mid &f_{k}\rangle =\mid 1\rangle _{Ak},  \label{Atqubits}
\end{eqnarray}%
we can rewrite (\ref{PsiA1Cbal}) and (\ref{PsiA1Cconst}) as 
\begin{equation}
|\psi \rangle _{A1-C}=\pm \frac{1}{\sqrt{2}}[(-1)^{F(0)}|0\rangle
_{A1}+(-1)^{F(1)}|1\rangle _{A1}]\frac{1}{\sqrt{2}}[|0\rangle
_{C}-(-1)^{F(0)\oplus F(1)}|1\rangle _{C}]
\end{equation}%
Now we make use of the Hadamard gate $H$ (\ref{H}). Then, in the case of $%
F(0)=F(1)$ the atom will be in the state $\pm \frac{1}{\sqrt{2}}(|0\rangle
_{A1}+|1\rangle _{A1})$ or $\pm \frac{1}{\sqrt{2}}(|e_{1}\rangle
+|f_{1}\rangle )$ and after we apply the $H$ gate we get $\pm |0\rangle
_{A1} $ or $|e_{1}\rangle $. In the case of $f(0)\neq f(1)$ the atom will be
in the state $\pm \frac{1}{\sqrt{2}}(|0\rangle _{A1}-|1\rangle _{A1})$ or $%
\pm \frac{1}{\sqrt{2}}(|e_{1}\rangle -|f_{1}\rangle )$ and after we apply
the $H$ gate we get $\pm |1\rangle _{A1}$ or $|f_{1}\rangle $. Notice that
the state $\frac{1}{\sqrt{2}}[|0\rangle -(-1)^{F(0)\oplus F(1)}|1\rangle ]$
in which the cavity is left is not important.

Another possible implementation of the Deutsch algorithm is possible
considering a three-level cascade atom \ $Ak$ with $\mid e_{k}\rangle ,\mid
f_{k}\rangle $ and $\mid g_{k}\rangle $ being the upper, intermediate and
lower atomic states. As above, we assume that the transition $\mid
f_{k}\rangle \rightleftharpoons \mid e_{k}\rangle $ is far enough from
resonance with the cavity central frequency such that only virtual
transitions occur between these states (only these states interact with
field in cavity $C$). In addition we assume that the transition $\mid
e_{k}\rangle \rightleftharpoons \mid g_{k}\rangle $ is highly detuned from
the cavity frequency so that there will be no coupling with the cavity field
(see Fig. 3). Here we are going to consider the effect of the atom-field
interaction taking into account only levels $\mid f_{k}\rangle $ and $\mid
g_{k}\rangle .$ We do not consider level $\mid e_{k}\rangle $ since it will
not play any role in our scheme. Therefore, we have effectively a two-level
system involving states $\mid f_{k}\rangle $ and $|g_{k}\rangle $.
Considering levels $\mid f_{k}\rangle $ and $\mid g_{k}\rangle ,$ we can
write an effective time evolution operator (see (\ref{U1})), 
\begin{equation}
U_{k}(t)=e^{i\varphi a^{\dagger }a}\mid f_{k}\rangle \langle f_{k}\mid
+|g_{k}\rangle \langle g_{k}\mid .  \label{U2}
\end{equation}%
A coherent state $|\alpha \rangle $ is obtained applying the displacement
operator $D(\alpha )=e^{(\alpha a^{\dag }-\alpha ^{\ast }a)}$ to the vacuum,
that is, $|\alpha \rangle =D(\alpha )|0\rangle ,$ and is given by 
\begin{equation}
|\alpha \rangle =e^{-\frac{1}{2}|\alpha |^{2}}{\sum\limits_{n=0}^{\infty }}%
\frac{(\alpha )^{n}}{\sqrt{n!}}|n\rangle .
\end{equation}%
We define the even and odd coherent states as 
\begin{eqnarray}
|+\rangle &=&\frac{1}{\sqrt{N^{+}}}(|\alpha \rangle +|-\alpha \rangle
)\equiv \frac{1}{\sqrt{2}}(|0\rangle _{C}+|1\rangle _{C}),  \nonumber \\
|-\rangle &=&\frac{1}{\sqrt{N^{-}}}(|\alpha \rangle -|-\alpha \rangle
)\equiv \frac{1}{\sqrt{2}}(|0\rangle _{C}-|1\rangle _{C}),  \label{EOCS}
\end{eqnarray}%
with $N^{\pm }=2\left( 1\pm e^{-2\mid \alpha \mid ^{2}}\right) \cong 2$ \
and $\langle +\mid -\rangle =0.$ \ Now, let us see how we can prepare a
state $|-\rangle $. Suppose we prepare cavity $C$ initially in a coherent
state $|-\alpha \rangle $. Then we prepare a two-level atom $B$, with $\mid
f\rangle $ and $\mid g\rangle $ being the upper and lower state
respectively, in a coherent superposition, \ sending $B$ in the lower state $%
\mid g\rangle $ through a first Ramsey cavity $R1$ where the atomic states
are rotated according to%
\begin{equation}
R_{1}=\frac{1}{\sqrt{2}}\left[ 
\begin{array}{cc}
c_{g} & c_{f} \\ 
-c_{f} & c_{g}%
\end{array}%
\right] ,
\end{equation}%
and we get%
\begin{equation}
\mid \psi \rangle _{B}=c_{f}\mid f\rangle +c_{g}\mid g\rangle .
\end{equation}%
After that, $B$ flies through cavity $C$ and, taking into account the time
evolution operator (\ref{U2}), after $B$ pass through $C$ the state of the
system $B-C$, for $\varphi =\pi $, is given by 
\[
\mid \psi \rangle _{B-C}=c_{f}\mid f\rangle \mid \alpha \rangle +c_{g}\mid
g\rangle \mid -\alpha \rangle 
\]%
Then, we send $B$ through a second Ramsey zone $R2$ \ where the atomic
states are rotated according to 
\begin{equation}
R_{2}=\frac{1}{\sqrt{2}}\left[ 
\begin{array}{cc}
1 & -i \\ 
-i & 1%
\end{array}%
\right] ,
\end{equation}%
that is,%
\begin{eqnarray}
&\mid &f\rangle \rightarrow \frac{1}{\sqrt{2}}(\mid f\rangle -i\mid g\rangle
),  \nonumber \\
&\mid &g\rangle \rightarrow \frac{1}{\sqrt{2}}(-i\mid f\rangle +\mid
g\rangle ),
\end{eqnarray}%
and therefore, the state of the system $B-C$ will be 
\begin{eqnarray*}
|\psi \rangle _{B-C} &=&\frac{1}{\sqrt{2}}[(c_{f}-ic_{g})\mid +\rangle
+(c_{f}+ic_{g})\mid -\rangle ]\mid f\rangle \\
&&+\frac{1}{\sqrt{2}}[(-ic_{f}+c_{g})\mid +\rangle )-(ic_{f}+c_{g})\mid
-\rangle ]\mid g\rangle ,
\end{eqnarray*}%
Now, in order to obtain a state $|-\rangle $ in cavity $C$, we detect atom $%
B $ in $\mid f\rangle $ or in $\mid g\rangle $. If we detect $\mid f\rangle $
we must have have\ $c_{f}=ic_{g}$ \ and if we detect $\mid g\rangle $ we we
must have have\ $c_{f}=-ic_{g}.$

Then, we start assuming that the cavity is prepared in a state $|-\rangle $
and atom $A1$ is prepared in a state%
\begin{equation}
\mid \psi \rangle _{A1}=\frac{1}{\sqrt{2}}(\mid f_{1}\rangle +\mid
g_{1}\rangle ).  \label{Psiat2}
\end{equation}%
After the atom fly through the cavity, for $\varphi =\pi $, we have%
\begin{equation}
|\psi \rangle _{A1-C}=\frac{1}{2}(-|f_{1}\rangle +|g_{1}\rangle )(|0\rangle
_{C}-|1\rangle _{C}).  \label{PsiA1Cbal2}
\end{equation}%
If $\varphi =2\pi $ we have 
\begin{equation}
|\psi \rangle _{A1-C}=\frac{1}{2}(|f_{1}\rangle +|g_{1}\rangle )(|0\rangle
_{C}-|1\rangle _{C}).  \label{PsiA1Cconst2}
\end{equation}%
Now, if we use the notation

\begin{eqnarray}
&\mid &f_{k}\rangle =\mid 0\rangle _{Ak},  \nonumber \\
&\mid &g_{k}\rangle =\mid 1\rangle _{Ak},  \label{Atqubits2}
\end{eqnarray}%
we can rewrite (\ref{PsiA1Cbal2}) and (\ref{PsiA1Cconst2}) as 
\begin{equation}
|\psi \rangle _{A1-C}=\pm \frac{1}{\sqrt{2}}[(-1)^{F(0)}|0\rangle
_{A1}+(-1)^{F(1)}|1\rangle _{A1}]\frac{1}{\sqrt{2}}[|0\rangle _{C}-|1\rangle
_{C}]
\end{equation}%
Then, in the case of $F(0)=F(1)$ the atom will be in the state $\pm \frac{1}{%
\sqrt{2}}(|0\rangle _{A1}+|1\rangle _{A1})$ or $\pm \frac{1}{\sqrt{2}}%
(|f_{1}\rangle +|g_{1}\rangle )$ and after we apply the $H$ gate we get $\pm
|0\rangle _{A1}$ or $|f_{1}\rangle $. In the case of $F(0)\neq F(1)$ the
atom will be in the state $\pm \frac{1}{\sqrt{2}}(|0\rangle _{A1}-|1\rangle
_{A1})$ or $\pm \frac{1}{\sqrt{2}}(|f_{1}\rangle -|g_{1}\rangle )$ and after
we apply the $H$ gate we get $\pm |1\rangle _{A1}$ or $|g_{1}\rangle $.

The speed up achieved in the Deutsch algorithm over the classical algorithm
is just a factor of two. However, in a generalized problem, we are going to
discuss now, the speedup is much more than in the classical case showing the
power of the quantum computation. Let us examine now the Deutsch-Jozsa
algorithm to solve the generalized Deutsch problem. Let $F:\{0,1\}^{\otimes
n}\longrightarrow \{0,1\}$ be a Boolean function of a n-bit integer argument
and we assume allow only those $f$ that are either constant or yield 0 for
exactly half of the arguments and 1 for the rest, that is, it is balanced.
Given an oracle that evaluates the function for a given argument, the
problem is to decide if $F$ is constant or balanced. There are $2^{n}$
possible arguments and to solve the problem classically we will have to
calculate the function for $2^{n-1}+1$ arguments in the worst case. Then, we
see that the computational resources required to solve the problem grows
exponentially with the bit size $n$ of the input. The Deutsch-Jozsa
algorithm can however solve this problem very easily. This algorithm makes
use of a quantum $f-$gate that is a generalization of the one we used above
in the Deutsch algorithm. The action in the computational basis is similar
to that of the ordinary $F-$gate, except that $|X\rangle $ here is a
computational basis state of a $n-$bit register. If the bottom qubit is in
the state $\frac{1}{\sqrt{2}}(|0\rangle -|1\rangle )$, then the state of the
upper register is transformed according to%
\begin{equation}
|X\rangle \longrightarrow (-1)^{F(X)}|X\rangle .
\end{equation}%
The quantum circuit of the Deutsch-Jozsa algorithm is shown in Fig. 4. The
upper input is $|0\rangle |0\rangle ...|0\rangle $ (with $n$ factors) and $%
H^{\otimes n}=H\otimes H\otimes ...\otimes H$. Then, the effect of the first
Hadamard gates on the top input state is%
\begin{equation}
H^{\otimes n}|0\rangle |0\rangle ...|0\rangle =\frac{1}{\sqrt{2^{n}}}{{\sum }%
_{X=0}^{2^{n}-1}}|X\rangle .
\end{equation}%
The Hadamard gate acting on the bottom input state $|1\rangle $ produces $%
\frac{1}{\sqrt{2}}(|0\rangle -|1\rangle )$. Thus the $f-$gate changes the
state of the register to%
\begin{equation}
\frac{1}{\sqrt{2^{n}}}{{{\sum }_{X=0}^{2^{n}-1}}(-1)^{F(X)}}|X\rangle .
\end{equation}%
Now notice that the action of the Hadamard gates on a computational basis
state is given by%
\begin{equation}
H^{\otimes n}|X\rangle =\frac{1}{\sqrt{2^{n}}}{{{\sum }_{Y=0}^{2^{n}-1}}%
(-1)^{X.Y}}|Y\rangle .
\end{equation}%
where $X.Y$ is the bitwise scalar product where for $X=x_{n-1}...x_{1}x_{0}$
and $Y=y_{n-1}...y_{1}y_{0}$ we have $X.Y=\oplus _{i=0}^{n-1}x_{i}y_{i}$ and 
$\oplus $ is addition mod 2. \ Then, the state of the register will be 
\begin{equation}
\frac{1}{\sqrt{2^{n}}}{{\sum }_{X=0}^{2^{n}-1}}{(-1)^{f(X)}}H^{\otimes
n}|X\rangle =\frac{1}{2^{n}}{{\sum }_{X=0}^{2^{n}-1}}{{{\sum }%
_{Y=0}^{2^{n}-1}}(-1)^{F(X)+X.Y}}|Y\rangle .
\end{equation}%
Now, the amplitude of $|Y=0\rangle =|0\rangle |0\rangle ...|0\rangle $ is $%
\frac{1}{2^{n}}{{\sum }_{X=0}^{2^{n}-1}}{(-1)^{f(X)}}$ and if $F$ is
constant this gives us $\pm 1$. On the other hand, if $f$ is balanced one
half of the terms in the sum cancels exactly the other half and the result
is 0. Then, the probability of observing $|Y=0\rangle =|0\rangle |0\rangle
...|0\rangle $ is 1 if $F$ is constant and is 0 if it is balanced.

Let us see now how we can implement the Deutsch-Jozsa algorithm
experimentally. Considering atoms $A1,A2,...,An$ prepared in a state like (%
\ref{Psiat}). After they fly through $C$ prepared in state (\ref{C-}),
taking into account (\ref{U1}) for $\varphi =\pi $ we have%
\begin{equation}
|\psi \rangle _{A1-C}=\frac{1}{\sqrt{2}}(-|e_{1}\rangle +|f_{1}\rangle )%
\frac{1}{\sqrt{2}}(-|e_{2}\rangle +|f_{2}\rangle )...\frac{1}{\sqrt{2}}%
(-|e_{n}\rangle +|f_{n}\rangle )\frac{1}{\sqrt{2}}(|0\rangle
_{C}-(-1)^{n}|1\rangle _{C}).  \label{PsiA1A2AnCbal}
\end{equation}%
If $\varphi =2\pi $ we have 
\begin{equation}
|\psi \rangle _{A1-C}=\frac{1}{\sqrt{2}}(|e_{1}\rangle +|f_{1}\rangle )\frac{%
1}{\sqrt{2}}(|e_{2}\rangle +|f_{2}\rangle )...\frac{1}{\sqrt{2}}%
(|e_{n}\rangle +|f_{n}\rangle )\frac{1}{\sqrt{2}}(|0\rangle _{C}-|1\rangle
_{C}).  \label{PsiA1A2AnCconst}
\end{equation}

If we assume atoms $A1,A2,...,An$ prepared in a state like (\ref{Psiat2})
and cavity $C$ prepared in state $|-\rangle $, taking into account (\ref{U2}%
) for $\varphi =\pi $ we have%
\begin{equation}
|\psi \rangle _{A1-C}=\frac{1}{\sqrt{2}}(-|f_{1}\rangle +|g_{1}\rangle )%
\frac{1}{\sqrt{2}}(-|f_{2}\rangle +|g_{2}\rangle )...\frac{1}{\sqrt{2}}%
(-|f_{n}\rangle +|g_{n}\rangle )\frac{1}{\sqrt{2}}(|0\rangle _{C}-|1\rangle
_{C}).  \label{PsiA1A2AnCbal2}
\end{equation}%
If $\varphi =2\pi $ we have 
\begin{equation}
|\psi \rangle _{A1-C}=\frac{1}{\sqrt{2}}(|f_{1}\rangle +|g_{1}\rangle )\frac{%
1}{\sqrt{2}}(|f_{2}\rangle +|g_{2}\rangle )...\frac{1}{\sqrt{2}}%
(|f_{n}\rangle +|g_{n}\rangle )\frac{1}{\sqrt{2}}(|0\rangle _{C}-|1\rangle
_{C}).  \label{PsiA1A2AnCconst2}
\end{equation}

Now, if we use the notation (\ref{Atqubits}) and (\ref{Atqubits2}) we can
rewrite (\ref{PsiA1A2AnCbal}), (\ref{PsiA1A2AnCconst}), (\ref{PsiA1A2AnCbal2}%
) and (\ref{PsiA1A2AnCconst2}) as 
\begin{equation}
|\psi \rangle _{A1-C}=\pm \frac{1}{\sqrt{2^{n}}}{{{\sum }_{X=0}^{2^{n}-1}}%
(-1)^{F(X)}}|X\rangle _{A}|\psi \rangle _{C}
\end{equation}%
where, in the case of (\ref{PsiA1A2AnCbal}) and (\ref{PsiA1A2AnCconst}), {if 
}$F$ {is balanced,} 
\begin{equation}
|\psi \rangle _{C}=\frac{1}{\sqrt{2}}(|0\rangle _{C}-(-1)^{n}|1\rangle _{C}){%
,}
\end{equation}%
and {if }$f$ {is constant,} 
\begin{equation}
|\psi \rangle _{C}=\frac{1}{\sqrt{2}}(|0\rangle _{C}-|1\rangle _{C}){.}
\end{equation}%
In the case of (\ref{PsiA1A2AnCbal2}) and (\ref{PsiA1A2AnCconst2}), 
\begin{equation}
|\psi \rangle _{C}=\frac{1}{\sqrt{2}}(|\alpha \rangle -|-\alpha \rangle
)\equiv \frac{1}{\sqrt{2}}(|0\rangle _{C}-|1\rangle _{C}){.}
\end{equation}

Now notice that the action of the $H$ gates on a computational basis state
is given by%
\begin{equation}
H^{\otimes n}|X\rangle _{A}=\frac{1}{\sqrt{2^{n}}}{{{\sum }_{Y=0}^{2^{n}-1}}%
(-1)^{X.Y}}|Y\rangle _{A}.
\end{equation}%
Then, the state of the register will be 
\begin{equation}
\frac{1}{\sqrt{2^{n}}}{{\sum }_{X=0}^{2^{n}-1}}{(-1)^{F(X)}}H^{\otimes
n}|X\rangle _{A}=\frac{1}{2^{n}}{{\sum }_{X=0}^{2^{n}-1}}{{{\sum }%
_{Y=0}^{2^{n}-1}}(-1)^{F(X)+X.Y}}|Y\rangle _{A}.
\end{equation}%
Now, the amplitude of $|0\rangle _{A}=|0\rangle _{A1}|0\rangle
_{A2}...|0\rangle _{An}$ is $\frac{1}{2^{n}}{{\sum }_{X=0}^{2^{n}-1}}{%
(-1)^{F(X)}}$ and if $F$ is constant this gives us $\pm 1$. On the other
hand, if $f$ is balanced one half of the terms in the sum cancels exactly
the other half and the result is 0. Then, the probability of observing $%
|0\rangle _{A}=|0\rangle _{A1}|0\rangle _{A2}...|0\rangle _{An}$ is 1 if $F$
\ is constant and is 0 if it is balanced. Therefore, if we detect $|0\rangle
_{A1}|0\rangle _{A2}...|0\rangle _{An}$ then $F$ is constant. \ Concluding,
a single call to the quantum oracle followed by a measure of the register
and checking if the result is $|0\rangle _{A1}|0\rangle _{A2}...|0\rangle
_{An}$ allows us to decide if the function is constant or balanced and we
have achieved an exponential speedup over the classical computation.
Finally, let us analyze the feasibility of the experimental implementation
of the above algorithms. Considering Rydberg atoms of principal quantum
numbers 50 or 51, the radiative time is of the order of $10^{-2}$ s and the
coupling constant $g$ is of the order of $2\pi \times 25$ kHz \cite{Rat1,
Rat2, Rat3} and the detuning. $\Delta $ is of the order of $2\pi \times 100$
kHz. Taking into account that $\varphi =g^{2}\tau /\Delta $, for $\varphi
=\pi $ we have an interaction time $\tau =8\times 10^{-5}$ s and we could,
in principle, assume a time of the order of $10^{-4}$ s to realize the
algorithm which is much shorter than the radiative time. We have to consider
also the cavity decay time which in recent experiments, with niobium
superconducting cavities at very low temperature and quality factors in the $%
10^{9}-10^{10}$ \ range, have a cavity energy damping time of the order of $%
10$ to $100$ ms, \ and which could be larger than the required time to
perform the algorithm.

\textbf{Figure Captions} \newline

\textbf{Fig. 1 -} Quantum circuit of the Deutsch algorithm.\newline

\textbf{Fig. 2 - }Energy states scheme of a two-level atom where $|e\rangle $
is the upper state with atomic frequency $\omega _{e}$, $\ |f\rangle $ is
the intermediate state with atomic frequency $\omega _{f},$ and $\omega $ is
the cavity field frequency and $\Delta =(\omega _{e}-\omega _{f})-\omega $
is the detuning. The transition $\mid f\rangle \rightleftharpoons \mid
e\rangle $ is far enough of resonance with the cavity central frequency such
that only virtual transitions occur between these levels.\bigskip

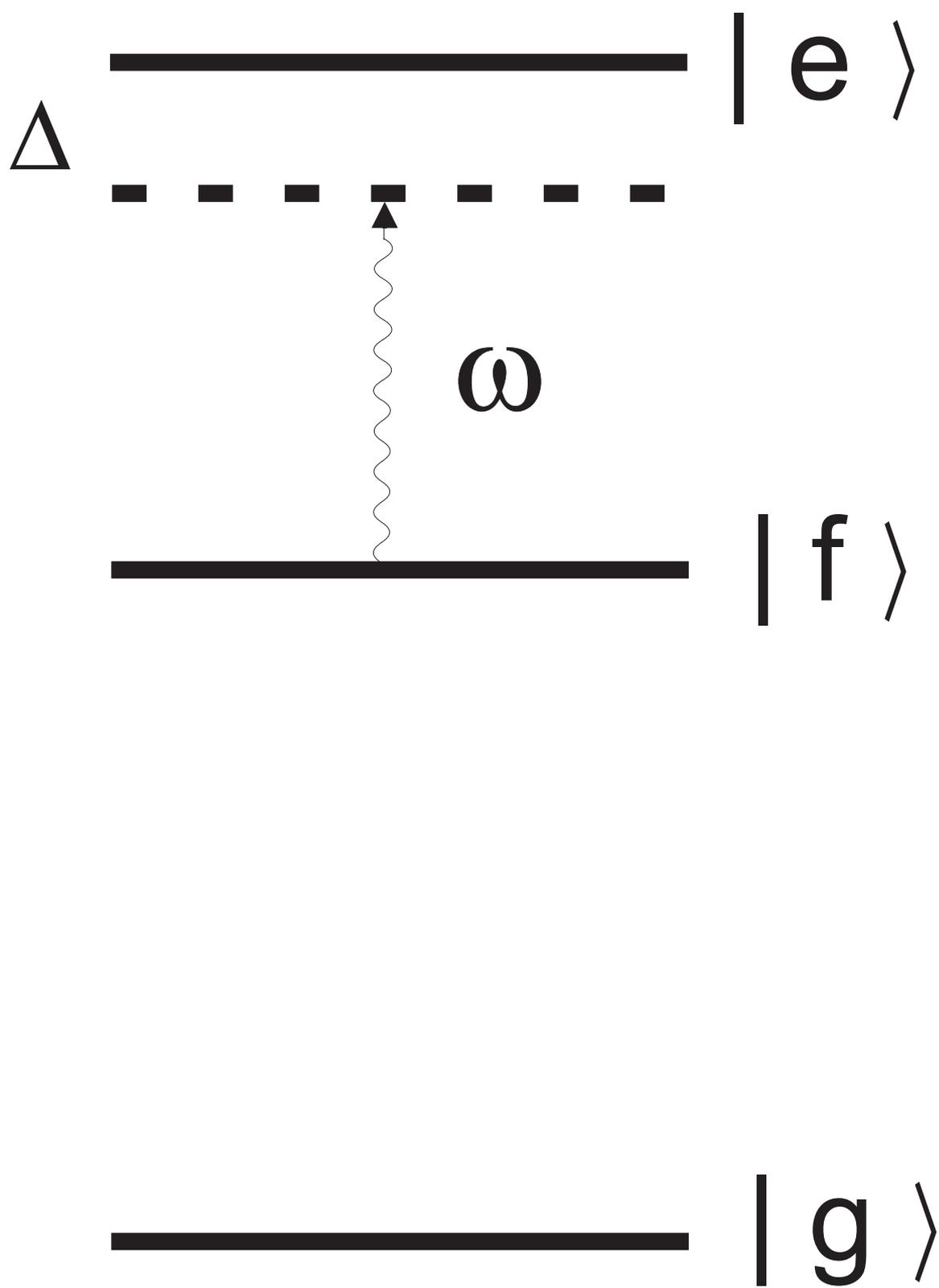
\textbf{Fig. 3} - Energy states scheme of a three-level atom where $%
|e\rangle $ is the upper state with atomic frequency $\omega _{e}$, $\
|f\rangle $ is the intermediate state with atomic frequency $\omega _{f}$, $%
|g\rangle $ is the lower state with atomic frequency $\omega _{g}$ and $%
\omega $ is the cavity field frequency and $\Delta =(\omega _{e}-\omega
_{f})-\omega $ is the detuning. The transition $\mid f\rangle
\rightleftharpoons \mid e\rangle $ is far enough of resonance with the
cavity central frequency such that only virtual transitions occur between
these levels (only these states interact with field in cavity $C$). In
addition we assume that the transition $\mid e\rangle \rightleftharpoons
\mid g\rangle $ is highly detuned from the cavity frequency so that there
will be no coupling with the cavity field in $C$.\bigskip

\textbf{Fig. 4 - }Quantum circuit of the Deutsch-Jozsa algorithm.

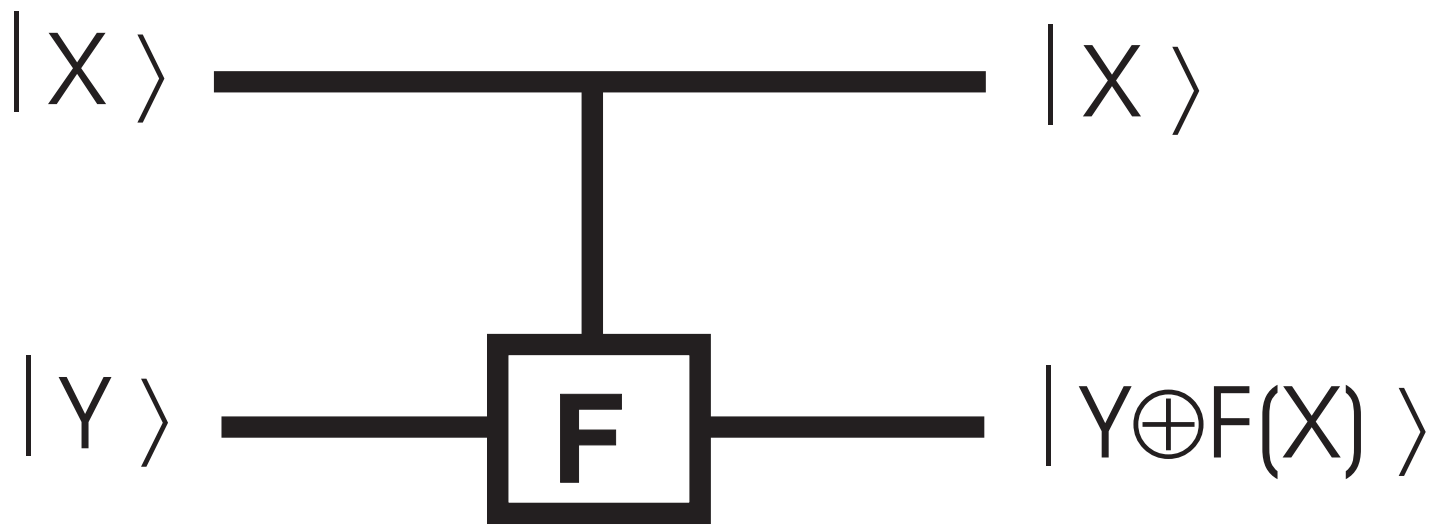
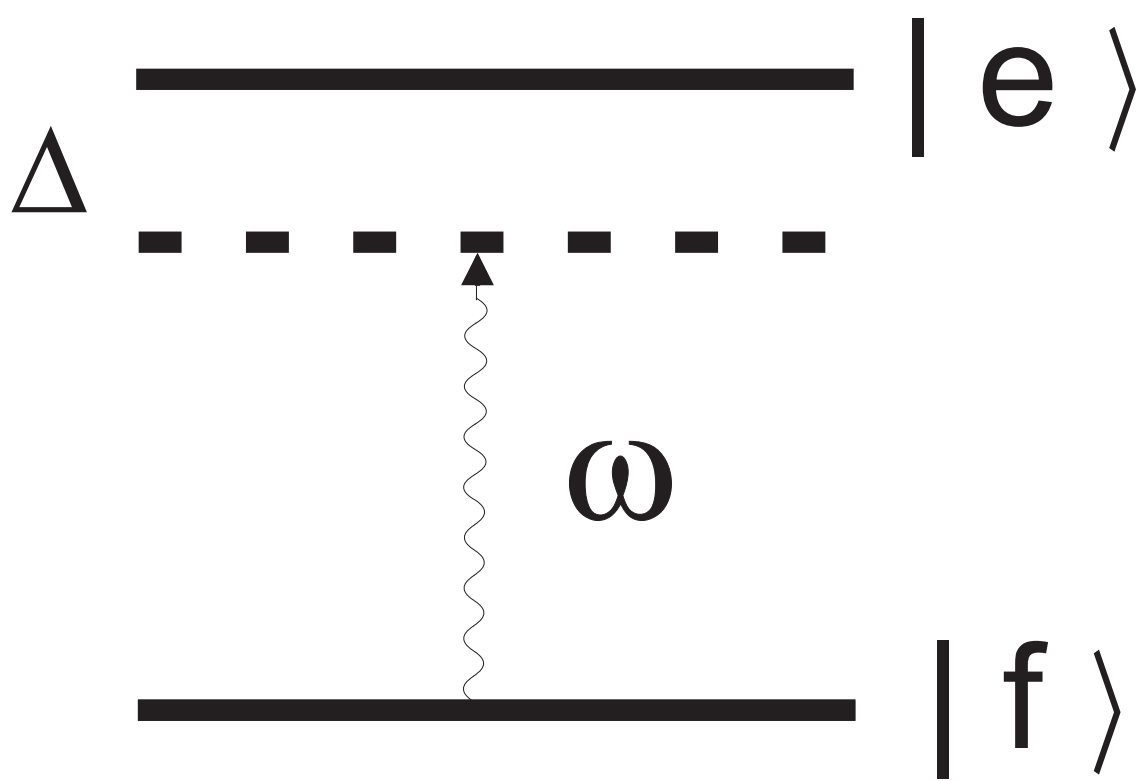
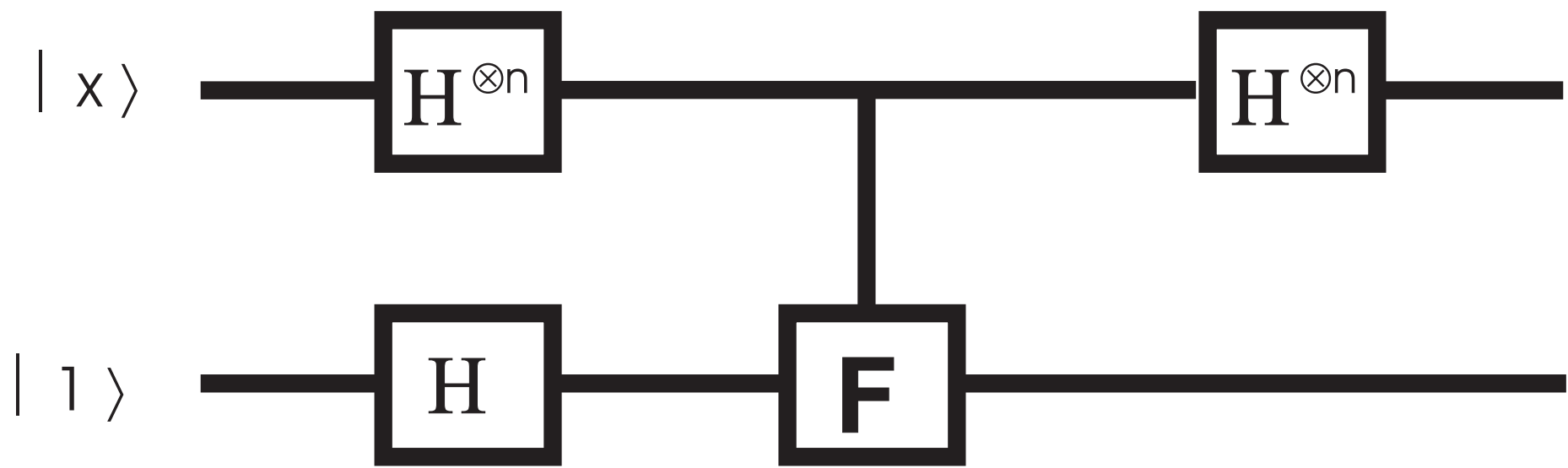
\end{document}